\documentclass[twocolumn,showpacs,preprintnumbers,amsmath,amssymb]{revtex4}

\usepackage{graphicx}
\usepackage{bm}
\usepackage{amsmath}
\usepackage{amsfonts}

\begin{document}
\title{Stable Hopf solitons in rotating Bose-Einstein condensates}
\date{\today}
\author{Y.M. Bidasyuk$^{1}$, A.V. Chumachenko$^2$, O.O. Prikhodko$^2$,  S.I. Vilchinskii$^2$, M. Weyrauch$^{1}$, A.I. Yakimenko$^2$}
\affiliation{
$^1$Physikalisch-Technische Bundesanstalt, Bundesallee 100, D-38116 Braunschweig, Germany \\
$^2$Department of Physics, Taras Shevchenko National University of Kyiv, Volodymyrska Str. 64/13, Kyiv 01601, Ukraine}
\begin{abstract}
 We reveal that Hopf solitons can be stabilized in rotating atomic Bose-Einstein condensates. The Hopfion is a matter-wave vortex complex
 which carries two independent winding numbers. Such a topological solitonic structure results from a superfluid flow of atoms simultaneously quantized in poloidal and toroidal directions. In the framework of a dissipative mean-field model we observe different unstable evolution scenarios of the Hopfions. We demonstrate energetic and dynamical stability of the Hopf solitons under experimentally feasible conditions.

\end{abstract}

\maketitle

\section{Introduction}

Quantization of vorticity is a remarkable manifestation of superfluid properties of Bose-Einstein condensates (BECs).
Different kinds of vortex structures have been theoretically predicted and observed experimentally in atomic BECs  (see e.g. \cite{Kevrekidis08,anderson2010resource,fetter2001vortices,bagnato2015bose} and references therein): single vortex lines, vortex-antivortex pairs, vortex arrays, vortex rings, vortex knots, dimerons \cite{PhysRevA.91.043616}, three-dimensional vortex solitons \cite{malomed2007stability}, and solitary waves moving along a straight vortex line \cite{Berloff05} (which are similar to ``hoop'' structures known in field theory \cite{RaduVolkovPhysRep08}).

It is of special interest to construct stable vortex-soliton states with complex structure, such as vortex knots \cite{PhysRevLett.100.180403,2010NatPh,2013NatPh}, three-dimensional Skyrmions, \cite{PhysRevLett.91.010403,PhysRevA.72.043616,PhysRevLett.86.3934,PhysRevA.70.041601} and Hopfions.  A Hopfion (or Hopf soliton) is a topological soliton with two independent winding numbers: the first, $S$, characterizes a horizontal circular vortex embedded into a three-dimensional soliton; and the second, $m$, corresponds to
vorticity around the axis, perpendicular to this circle.  Hopf solitons appear in many fields, including field theory, optics, ferromagnets, and semi- and superconductors. Hopfions have been predicted theoretically both in systems with multicomponent wave functions \cite{PhysRevLett.81.4798} and in single-component BEC \cite{PhysRevLett.113.264101}.
Multicharged ($m>1$, $S>1$) vortex structures have been demonstrated ~\cite{Berry} to be unstable in optical media. Stationary Hopfion solutions with $m=1$, $S=1$ of the single-component Gross-Pitaevskii-Equation (GPE) have been obtained numerically in Ref. \cite{Kao07} for a spherically-symmetric trapping potential. To the best of our knowledge, neither a dynamical nor an energetic stability analysis has been performed previously for  Hopfions in trapped BECs. Very recently \cite{PhysRevLett.113.264101}
Hopfions  have been demonstrated to be stable in a single-component BEC without linear trapping potential but with  a repulsive non-linearity, which grows fast enough from the center to the periphery of the trap.

In our recent work \cite{PhysRevA.88.043637} we suggested  an experimentally feasible trapping configuration that can be used to create, stabilize, and
manipulate a vortex ring in a controllable and nondestructive manner. In this paper we use a similar trapping potential and a rotating condensate which allows the formation of a stable Hopfion. We demonstrate both energetic and dynamical stability of the
Hopf soliton for realistic experimental parameters.

\section{Model}

The properties of an atomic BEC close to thermodynamic equilibrium can be accurately described in mean-field approximation by a dissipative Gross-Pitaevskii  equation (DGPE), which takes the following form in a rotating frame \cite{RevModPhys.81.647,pitaevskii1958phenomenological,choi1998phenomenological},
\begin{eqnarray}\label{GPE}\nonumber
(i-\gamma) \hbar \frac{\partial \Psi(\textbf{r},t)}{\partial t} = \left[-\frac{\hbar^2}{2M} \nabla^2 +  V_\textrm{ext}(\textbf{r}) -  \Omega L_z \right.\\
\left. + g |\Psi(\textbf{r},t)|^2 -  \mu \right]\Psi(\textbf{r},t).
\end{eqnarray}
Here, $\gamma \ll 1$ is a phenomenological dissipation parameter, $\Omega$ the rotation rate,
$\nabla^2$ is the Laplace operator, $L_z = -i\hbar(x\partial_y - y\partial_x)$ is the angular momentum in $z$-direction, $ g = 4 \pi \hbar^2 a_s/{M}$ is the coupling strength, $\mu$ is the chemical potential of the condensate, $M$ is the mass of the atom, and  $a_s$ is the $s$-wave scattering length. We assume that the wave function is normalized to the number of atoms
\begin{equation}\label{N}
 N=\int|\Psi|^2d\textbf{r}.
\end{equation}
In the absence of dissipation ($\gamma = 0$), the particle number $N$ as well as the total energy
\begin{eqnarray}\label{E3D}\nonumber
E=\int\left\{ \frac{\hbar^2}{2M} |\nabla \Psi|^2  + V_\textrm{ext}(\textbf{r})|\Psi|^2  \right. \\
\left. - \Omega \mathrm{Re}(\Psi^* L_z \Psi)+\frac{ g}{2}|\Psi|^4\right\}d\textbf{r}
\end{eqnarray}
are integrals of motion.

In our model the external trapping potential
\begin{equation}\label{trapConfig}
V_\textrm{ext}(\textbf{r})=V_h(r,z)+V_b(r,z)
\end{equation}
is created by an oblate harmonic trap
$$V_h(r,z)=\frac12 M (\omega_z^2 z^2+\omega_r^2 r^2)$$
and two blue-detuned laser beams: a radial Laguerre-Gaussian beam LG$_{0}^q$
and an elliptic highly anisotropic ``sheet'' beam creating a tight repulsive potential in $z$-dimension:
$$
V_b(r,z)=v_q \left(\frac{r}{R_{\rm T}}\right)^{2q}e^{-q[(r/R_{\rm T})^2-1]} + v_z e^{-(z/Z_{\rm T})^2},
$$
 where
 $r=\sqrt{x^2+y^2}$,
 $Z_\textrm{T}$ is the effective width of the sheet-beam, $R_\textrm{T}$ is the radial coordinate of the trap minimum, and $q$ is the topological charge of the red-detuned LG$_{0}^q$ optical beam.

\begin{figure}[htb]
\includegraphics[width=\linewidth]{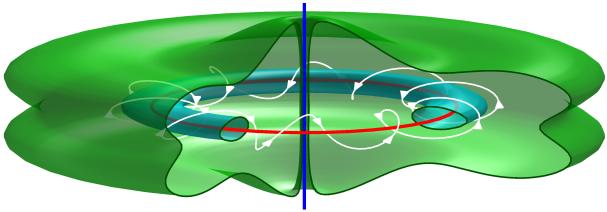}
\caption{(Color online) Schematic illustration of a vortex Hopfion in a trapped BEC. We show a condensate isodensity surface consisting of the outer surface (green) and the interior toroidal void (light blue). A Hopfion is formed by two types of phase singularities: the vertical vortex line (blue) surrounded by the vortex ring (red). The white line with arrows shows schematically the particle flow.
}\label{Torus}
\end{figure}

 Let us consider a BEC cloud  with $N= 10^6$ of $^{87}$Rb atoms ($a_s=5.77\,$nm) in an oblate harmonic trap with $\omega_r= 2 \pi \times 100$ Hz, $\omega_z=2 \pi \times 300$ Hz, and a characteristic oscillatory length $l_r=\sqrt{\hbar/(M\omega_r)}=2.7\,\mu$m. The tube beam has parameters $q=1$,  $R_\textrm{T}=13.5\,\mu$m, and $v_q/h=3.4\,$kHz, the sheet beam has parameters $Z_\textrm{T}=3.38\,\mu$m, and $v_z/h=1.8\,$kHz, where $h$ is the Planck's constant. This trap geometry is similar to that used in Ref.~\cite{PhysRevA.88.043637}  modified slightly to increase the stability of the vortex line by squeezing the trap in the $z$ direction and shifting the toroidal void position farther from the center. Such a trap configuration is supposed to naturally accommodate the complex core structure of a Hopfion (see Fig.~\ref{Torus}).

In what follows we use dimensionless units for the spatial coordinates $(x,y,z)\to (x/l_r,y/l_r,z/l_r)$, for time $t \to \omega_r t$, and for other quantities $\Psi \to  \Psi\sqrt{l_r^3}$,
 $E \to  E/(\hbar\omega_r)$, $\xi \to \xi/l_r$, and $\Omega \to \Omega/\omega_r$.
The DGPE then becomes
\begin{eqnarray}\label{GPEdl}\nonumber
(i-\gamma) \frac{\partial \Psi(\textbf{r},t)}{\partial t} = \left[-\frac{1}{2} \nabla^2 + V_\textrm{ext}(\textbf{r}) - \Omega L_z \right. \\
\left. + g |\Psi(\textbf{r},t)|^2 - \mu \right]\Psi(\textbf{r},t),
\end{eqnarray}
where $g=4\pi a_s/l_r = 2.76\times10^{-2}$.
The trapping potential in dimensionless units takes the form
$$
V_\textrm{ext}=\frac12 (\lambda^2 z^2+ r^2)+ v_q\beta^{2q}r^{2q}e^{-q(\beta^2 r^2-1)}+ v_z e^{-\alpha^2 z^2},
$$
with the  parameters
$\alpha=l_z/Z_\textrm{T}=0.8$, $\beta=l_z/R_{\textrm{T}}=0.2$,  $v_q = 34$, $v_z=18$, $\lambda = \omega_z/\omega_r = 3$.

For dynamical calculations the phenomenological dissipation parameter $\gamma=0.02$ was chosen to be constant, though general results do not qualitatively depend on this value.

\section{Energetic stability of vortex Hopfions}

If we aim to study the dependence of the total energy (\ref{E3D}) on the rotation rate $\Omega$ then it is convenient to split it into two parts
\begin{equation}\label{energyOmega}
E(\Omega) = E_0 - \Omega \mathcal{L}.
\end{equation}
The first part $E_0$ is the energy in the absence of rotation
$$
E_0 = \int \left[ \frac12 |\nabla \Psi|^2 + V_{ext} |\Psi|^2 + \frac g 2 |\Psi|^4\right]d\mathbf{r},
$$
and the second part $\mathcal{L}$ depends on the average of the angular momentum
$$
\mathcal{L}  = \int \mathrm{Re}(\Psi^* L_z \Psi) d\mathbf{r}.
$$

We obtain the BEC ground-state $\Psi_{\rm g.s.}$ by imaginary time propagation \cite{doi:10.1137/S1064827503422956} from Eq. (\ref{GPEdl}) with $\gamma=0$ and then compute its energy $E_{\rm GS}$ from Eq.~(\ref{E3D}).
Then, in order to model a Hopfion state characterized by topological charges $S$ and $m$, we imprint on the ground state an $S$-charged vortex ring with coordinates $(r_r,z_r)$ and an $m$-charged vortex line shifted by $r_l$ off the center.
\begin{equation}\label{psiImprint}
\Psi(\textbf{r}) =  \left\{f_r(\textbf{r}) e^{i \theta_r}\right\}^S \left\{f_l(\textbf{r})e^{i \theta_l}\right\}^m  \Psi_{\rm g.s.}(\textbf{r})
\end{equation}
with $\tan \theta_r=(z-z_r)/(r-r_r)$, $\tan \theta_l=y/(x-r_l)$, $r=\sqrt{x^2+y^2}$, $f_r(\textbf{r})=\tanh\left(\sqrt{(r-r_r)^2+(z-z_r)^2}/\xi\right)$, and $f_l(\textbf{r})=\tanh\left(\sqrt{(x-r_l)^2+y^2}/\xi\right)$. The width of the vortex core is estimated by the healing length $\xi$ at the peak density of the condensate.

If the vortex line is straight and parallel to the $z$-axis and the vortex ring is parallel to the $(x,y)$-plane, the superflows of the vortex ring and line are moving in perpendicular directions. Additionally, the density perturbations caused by the vortex cores are tightly localized and do not overlap in the regions of interest. Therefore, the contributions to the total energy from the vortex ring and vortex line are independent and can be analysed separately,
$$
E(\Omega) = E_{0,r}(r_r,z_r) + E_{0,l}(r_l) - \Omega \mathcal{L}_l (r_l),
$$
and we take into account that the vortex ring in the $(x,y)-$plane does not contribute to the angular momentum in $z$-direction.

\begin{figure}[htb]
\includegraphics[width=\linewidth]{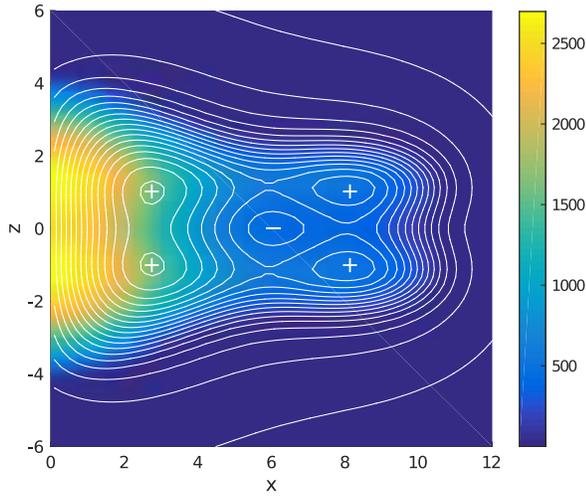}
\caption{(Color online) Density distribution of a trapped BEC cloud with contour lines of the vortex ring nucleation energy. Locations of energy extrema are denoted with ``plus'' (maxima) and ``minus'' (minimum) signs.}\label{energy_ring}
\end{figure}

\begin{figure}[htb]
\includegraphics[width=\linewidth]{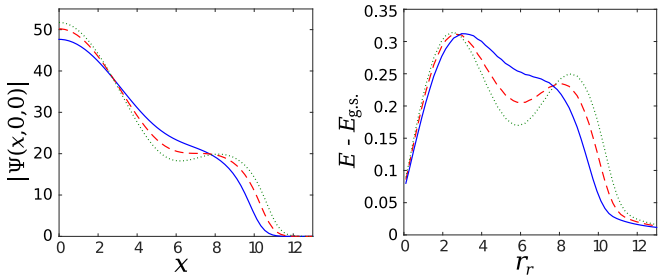}
\caption{(Color online) Absolute value of the wavefunction (left) and vortex ring nucleation energy (right) for different values of $v_q$ ($v_q=25$ --- blue solid line, $v_q=34$ --- red dashed line, $v_q=40$ --- green dotted line).}\label{energy_ring2}
\end{figure}

The energy analysis of the vortex ring is performed by putting $S=1$ and $m=0$ in (\ref{psiImprint}). Then the energy $E = E_{0,r}(r_r,z_r)$ depends only on $r_r$ and $z_r$ (see Fig.~\ref{energy_ring}).
 The vortex ring nucleation energy shows a minimum localized close to the density minimum at the crossing of the ``tube'' and ``sheet'' beams. It is worth noticing that the existence of an energy minimum does not necessarily correlate with the existence of a density minimum in the ground state. Figure~\ref{energy_ring2} shows that for certain intensities of the ``tube'' beam an off-center minimum exists in the vortex ring nucleation energy but not in the ground-state density.

The energy associated with the vortex line is obtained by putting $S=0$ and $m=1$ in Eq.~(\ref{psiImprint}). This energy $E(\Omega) = E_{0,l}(r_l) - \Omega \mathcal{L}_l (r_l)$ depends only on $\Omega$ and $r_l$. In analogy to the analysis of Fetter \cite{RevModPhys.81.647} we can define two characteristic rotation speeds. The first, $\Omega_c$, is chosen such that $E=E_{\rm g.s.}$ at $r_l=0$. It means that for $\Omega>\Omega_c$ a state with the vortex at the origin is energetically preferable over a state without vortex. This rotation rate can be defined as
$$
\Omega_c = \frac{E_{0,l}(0)-E_{\rm g.s.}}{\mathcal{L}_l (0)}.
$$
Note that $\mathcal{L}_l (0)/N = m = 1$. Therefore, in order to obtain this critical velocity only  $E_{0,l}(0)$ has to be calculated.

The second characteristic rotation rate $\Omega_m$ corresponds to a vanishing curvature of $E(r_l)$ at $r_l=0$:
$$
\Omega_m = \frac{\partial^2 E_{0,l}(r_l)}{\partial r_l^2} \bigg/ \frac{\partial^2 \mathcal{L}(r_l)}{\partial r_l^2}\bigg|_{r_l\to0}.
$$
This rotation rate defines the region $\Omega > \Omega_m$ where the energy shows a minimum at $r_l=0$ and the vortex line at the origin can be stable (or metastable). Due to our specific trap configuration there might be another energy minimum for the vortex line which is close to the peak intensity of the ``tube'' beam. But this minimum is not relevant for the stability of a Hopfion as it correlates with a stable position of the vortex ring. As the line and the ring have to be spatially separated, we demand the existence of a minimum at the origin as a situation favourable for a stable Hopfion.

\begin{figure}[htb]
\includegraphics[width=\linewidth]{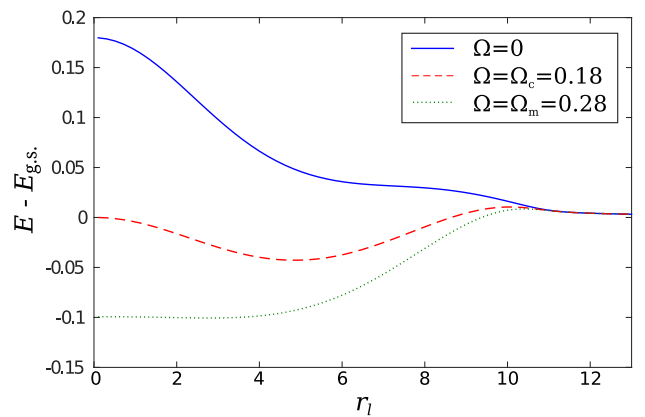}
\caption{(Color online) Nucleation energy of the vortex line depending on the displacement from the trap center for different values of the rotation rate.}\label{energy_line}
\end{figure}

As is seen from Fig.~\ref{energy_line}, in the suggested trap configuration, a vortex state can be energetically preferable without being stable at the center of the trap. This means that if we need to stabilize the vortex core at the axis, the rotation rate has to be considerably higher than $\Omega_c$. Therefore $\Omega_m > \Omega_c$, which is opposite to the case of a spherical condensate studied in Ref.~\cite{RevModPhys.81.647}. As a result we consider the obtained value $\Omega_m = 0.28$ as a minimal rotation rate at which the vortex line can be stable at the center of the trap and, therefore, at which we can also expect the overall stability of the Hopfion complex.

\section{Dynamics of Hopfions}

For the solution of the time-dependent GPE we use a split-step
Fourier transform method \cite{Agrawal}. The initial state for the simulations takes the same form Eq.~(\ref{psiImprint}) as in the energetic stability analysis of the previous section, i.e. a stationary state with imprinted vortex ring and line ($S=1$, $m=1$).
Their initial positions are considerably displaced from their equilibrium  (by up to $0.5\,l_r$).
We study the dynamical stability of this Hopfion solution for a wide range of rotation rates $\Omega$.

\begin{figure}[htb]
\includegraphics[width=\linewidth]{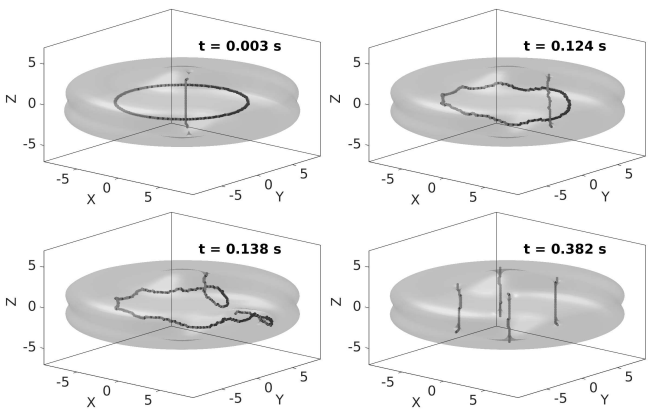}
\caption{Snapshots of the unstable Hopfion dynamics for the rotation rate $\Omega = 0.2$. In the initial state $S=1$, $m=1$. Reconnection happens at $t=0.135$ s. The four vortex lines in the final state are stable.}\label{unstable1_dynamics}
\end{figure}

We observe two different instability scenarios of the Hopfion dynamics. If the rotation rate is low or the initial position of the vortex line is too far away from equilibrium, the system is unstable with respect to reconnection of the line and ring cores. As a consequence, several line vortices are created which, in turn, can be stable and form a lattice at the off-center energy minimum (see Fig.~\ref{unstable1_dynamics}).

\begin{figure}[htb]
\includegraphics[width=\linewidth]{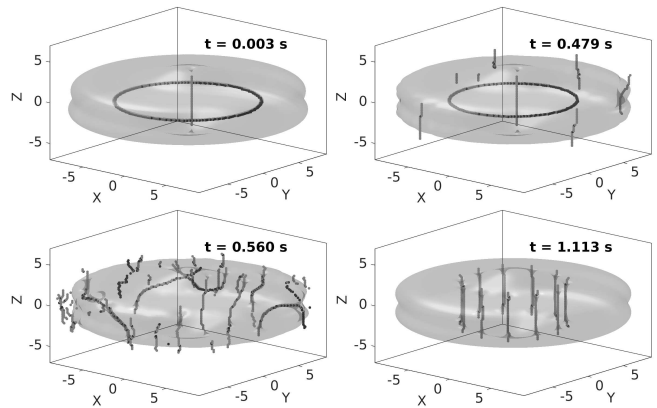}
\caption{Snapshots of the unstable Hopfion dynamics for the rotation rate $\Omega = 0.3$. In the initial state $S=1$, $m=1$. The Hopfion complex itself is stable, but starting from $t \approx 0.46$ s vortices from the outside start to penetrate the cloud eventually destroying the vortex ring.}\label{unstable2_dynamics}
\end{figure}

The second instability scenario is observed at high rotation rates and is independent of the initial displacement of the vortices. Due to a low energy barrier (see Fig.~\ref{energy_line}), additional vortices can be nucleated at the outer edge of the BEC cloud and then penetrate into the condensate destroying the ring vortex and again forming a regular vortex chain at the off-center energy minimum (see Fig.~\ref{unstable2_dynamics}).

\begin{figure}[htb]
\includegraphics[width=\linewidth]{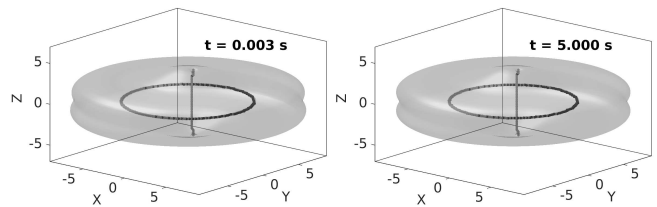}
\caption{Snapshots of the stable Hopfion dynamics for the rotation rate $\Omega = 0.24$. In the initial state $S=1$, $m=1$. Shapes of the vortex cores persist through all the time of simulation.}\label{stable_dynamics}
\end{figure}

It is however possible to choose a rotation rate that is high enough for stabilization of the vortex line at the center but at the same time low enough that additional vortices are not generated at the outer edge of the condensate. Then the vortex Hopfion may be dynamically stable for experimentally traceable time intervals which are comparable to the life-time of the condensate (see Fig.~\ref{stable_dynamics}). For a strong initial displacement of the vortex line of $0.5\,l_r$ we found a region of stability between $0.225 < \Omega < 0.245$. It is worth noticing that the vortex line relaxes to a position slightly off the center of the trap even though the energy minimum is located there. However, the energetic stability analysis does not take into account bending of the vortex cores, which are energetically  preferred over straight vortices. This may also explain the fact that the vortex line is dynamically stabilized at a rotation rate lower than predicted by the energetic analysis.

\begin{figure}[htb]
\includegraphics[width=\linewidth]{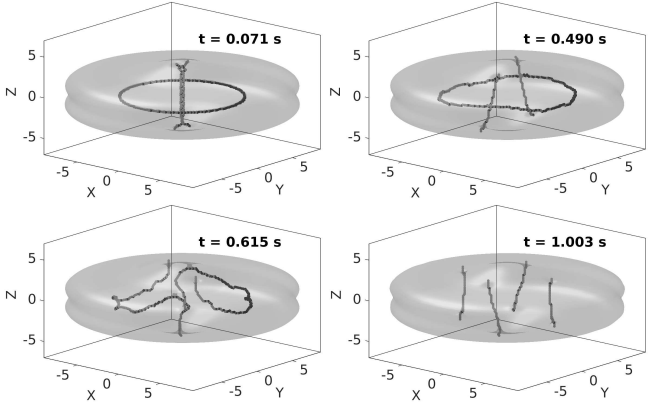}
\caption{Snapshots of the dynamics of the Hopfion with doubly-charged vortex line ($S=1$, $m=2$) for the rotation rate $\Omega = 0.25$.}\label{m2_dynamics}
\end{figure}

It is also instructive to analyze the dynamics of Hopfions with a topological charge of the vortex line $m > 1$. As multiply charged vortices in BEC are essentially unstable, the vortex line splits into a tangled complex of several singly charged vortices (see Fig.~\ref{m2_dynamics}). Because these vortices repel each other they can not all be simultaneously stabilized at the center of the trap. They rather relax to the energy minimum at the peak intensity of the ``tube'' beam. This inevitably leads to reconnection with the vortex ring and the observed dynamical scenario is similar to the instability scenario of singly charged vortex lines at low rotation rates (compare Figs.~\ref{unstable1_dynamics} and \ref{m2_dynamics}).

\begin{figure}[htb]
\includegraphics[width=\linewidth]{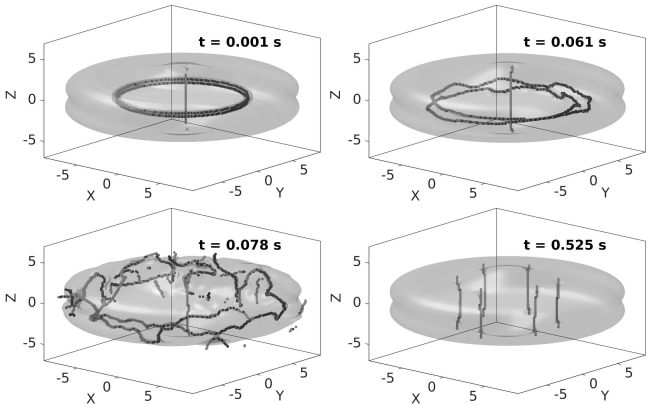}
\caption{Snapshots of the dynamics of the Hopfion with doubly-charged vortex ring ($S=2$, $m=1$) for the rotation rate $\Omega = 0.25$.}\label{s2_dynamics}
\end{figure}

Similarly, the Hopfion with a multicharged vortex ring ($S>1$) can not be stabilized in a suggested trap geometry (see Fig.~\ref{s2_dynamics}). 
The multi-charged vortex ring
splits into two leapfrogging singly-charged rings, that spiral around the energy minimum drifting away from it until they are destroyed by growing Kelvin waves. It is remarkable that in sharp contrast to studies of multicharged ($S>1$) Hopfions in optical media \cite{Berry} we do not observe the formation of vortex knots in our simulations of the dissipative dynamics of Hopf solitons in BEC.

The destruction of the Hopfion in the proposed setup always leads to formation of a vortex chain along the off-center energy minimum. The number of vortices in this chain is defined not only by the rotation rate but also by the total energy released by destruction of the Hopfion complex (note a six-vortex final state in Fig.~\ref{s2_dynamics}).

The main aim of the present work is a demonstration
that Hopfions can be stabilized in experimentally feasible
one-component trapped BECs. Suitable parameters sets for the stabilization of Hopfions depend on the specific experimental realization of BEC. We observed that Hopfions are more stable in highly oblate traps since vortex line bending and tilting are suppressed under these conditions \cite{PhysRevA.84.023637}. Furthermore,
imprinting the vortex line closer to the axis extends the stability region towards lower rotation rates.


\section{Conclusions}

We show that Hopf solitons can be  energetically and dynamically stable in a rotating trapped atomic BEC. In the framework of a dissipative mean-field model we investigate different dynamical regimes of Hopfion stability. It turns out that within a certain interval of BEC rotation velocities a Hopfion may be stable. For rotation rates lower than this interval the vortex line and ring deform and finally reconnect. For over-critical angular velocities, vortex lines are nucleated at the outer edge of the condensate and then reconnect with the vortex ring. Multiply charged vortex lines split into singly charged vortex lines, which then reconnect with the ring component.

Decay of the Hopfion leads to the formation of a vortex lattice.
The kinetic energy of the ring component is transferred via reconnection to the vortex  lattice. As is well known, the experimental detection of vortex rings is much more challenging than observation of the holes in the atomic cloud produced by vortex lines. By counting the number of vortex lines in the  vortex lattice one obtains clear experimental evidence of the presence of the vortex ring component.

We hope that the results presented in this paper will stimulate experiments to observe Hopf solitons in atomic BECs. Moreover, investigation of the Hopfion in the well controlled environment of atomic physics could help to elucidate the properties of these fascinating topological structures in other fields including superconductors, ferromagnets, and quantum field theory.

\begin{acknowledgments}
AC and AY thank the Physikalisch-Technische Bundesanstalt (PTB) for warm hospitality during their stay in Braunschweig.
AC also thanks the National University of Kiev for financial support given by the program ``100+100+100''.
AY acknowledges support from Project 1/30-2015  ``Dynamics and topological structures in Bose-Einstein condensates of ultracold gases''  of the KNU  Branch Target Training at the NAS of Ukraine.
S.V. is grateful for  the support of this work to
the  German Academic Exchange Service (DAAD),
 grant  No. 91563279 (Scholarship Programme" Research  Stays
for University Academics and Scientists, 2015")
and the Swiss National Science Foundation
grant SCOPE  IZ 7370-152581.
\end{acknowledgments}


\end{document}